\documentclass[aps,preprint,epsfig,rotate]{revtex4}
\begin{document}
\title{Hyperfine structure splitting of the positron-helium ions
       $e^{+}[^3$He$(2^3S)]$ and $e^{+}[^4$He$(2^3S)]$.}

 \author{Alexei M. Frolov}
 \email[E--mail address: ]{afrolov@uwo.ca}

\affiliation{Department of Chemistry\\
 University of Western Ontario, London, Ontario N6H 5B7, Canada}

\date{\today}

\begin{abstract}

The hyperfine structure splittings are determined for the lowest bound state 
in the positron-helium ion $e^{+}[^3$He$(2^3S)]$ and $e^{+}[^4$He$(2^3S)]$. 
In particular, we have fond that for the $e^{+}[{}^3$He$(2^3S)]$ ion one can 
observe the three following hyperfine structure splittinings: 5824.3986 $MHz$, 
76466.5308 $MHz$ and 5824.4019 $MHz$. In the $e^{+}[{}^4$He$(2^3S)]$ ion only 
one hyperfine structure splitting 82963.0427 $MHz$ can be observed. All these 
values can be measured in modern experiments. 

\end{abstract}

\maketitle
\newpage

The positron-helium ions are of some interest in Astrophysics and applications
related to the positron annihilation and positron conservation in light atomic
systems (see, e.g., \cite{Dra1}, \cite{Schra}, \cite{Klei}, \cite{Dra2} and 
references therein). In our earlier study \cite{Fro01} we have shown that the the 
positron-helium ion $e^{+}$He is bound, if (and only if) the two of its electrons 
are in the triplet state. In respect with this in \cite{Fro01} and in this study the 
bound positron-helium ion is designated as $e^{+}$[He$(2^3S)]$. The expectation 
values of various bound state properties in this ion were presented in \cite{Fro01}. 
Since then our computational results for the $e^{+}[$He$(2^3S)]$ ion have been 
improved substantially. However, in this study we want to investigate the hyperfine 
structure and evaluate the hyperfine structure splittings for the ground bound state 
of the $e^{+}$[He$(2^3S)]$ ion. This problem has never been solved accurately (see 
discussion in \cite{Fro01}). On the other hand, we have found that the hyperfine 
structures of the $e^{+}$[He$(2^3S)]$ ions are relatively reach and due to some 
reasons (see below) they are very interesting objects for investigation.
 
The operator responsible for the hyperfine structure splitting (or hyperfine 
splitting, for short) in the four-body $e^{+}[$He$(2^3S)]$ ion is written in the 
following form (in atomic units) (see, e.g., \cite{Fro01}, \cite{LLQ})
\begin{eqnarray}
 (\Delta H)_{h.s.} = \frac{2 \pi}{3} \alpha^2 \frac{g_{He} g_{-}}{m_p}
 \langle \delta({\bf r}_{He-e^{-}}) \rangle ({\bf I}_{He} \cdot {\bf S}_{-}) +
 \frac{2 \pi}{3} \alpha^2 g_{+} g_{-}
  \langle \delta({\bf r}_{+-}) \rangle ({\bf s}_{+} \cdot {\bf S}_{-}) 
 \nonumber \\
 + \frac{2 \pi}{3} \alpha^2 \frac{g_{He} g_{+}}{m_p}
  \langle \delta({\bf r}_{He-e^{+}}) \rangle ({\bf I}_{He} \cdot {\bf s}_{+})
 \label{e1}
\end{eqnarray}
where $\alpha = \frac{e^2}{\hbar c}$ is the fine structure constant, $m_p$ is the 
proton mass and $g_{He}, g_{-}$ and $g_{+}$ are the $g-$factors of the He-nucleus,
electron and positron, respectively. In this equation ${\bf S}_{-}$ is the total
vector of the two-electron spin, ${\bf I}_{He}$ is the spin of the nucleus and ${\bf 
s}_{+}$ is the positron spin. Note that the expression, Eq.(\ref{e1}), for
$(\Delta H)_{h.s.}$ is, in fact, an operator in the total spin space which has 
the dimension $N = (2 S_{-} + 1) \cdot 2 \cdot (2 I_{He} + 1) = 6 (2 I_{He} + 1)$. 
In the case of the ${}^3$He nucleus the dimension $N$ equals 12, while for the 
${}^4$He nucleus such a dimension ($N$) equals 6.

In our calculations we have used the following numerical values for the constants 
and factors in Eq.(\ref{e1}): $\alpha = 7.297352586 \cdot 10^{-3}, m_p = 1836.152701 
m_e, g_{-}$ = -2.0023193043622 and $g_{+} = - g_{-}$. The $g-$factor of the helium-3 
nucleus is deteremined from the formula: $g_{N} = \frac{{\cal M}_{N}}{I_{N}}$ = 
-4.2555016, where ${\cal M}_{N}$ = -2.1277508 \cite{CRC} is the magnetic moments (in 
nuclear magnetons) of the helium-3 nucleus. The spin of the helium-3 nucleus is 
$I_{He} = \frac12$. The both spin and $g-$factor of the helium-4 nucleus equal zero.  

The diagonalization of the matrix of the $(\Delta H)_{h.s.}$ operator, Eq.(\ref{e1}), 
leads to the conclusion that twelve spin states of the hyperfine structure of the 
$e^{+}[{}^3$He$(2^3S)]$ ion are separated into four different gourps which correspond 
to the following values of the total angular momentum $J$ = 1, 2, 0 and 1, respectively. 
The total number of hyperfine states in each group equals $2 J + 1$. The corresponding 
energies of these groups of states can be found in Table I. All these energies are 
expressed in $MHz$. The differences $\Delta_{J;J-1}$ between the corresponding hyperfine 
energies, i.e. the values
\begin{equation}
 \Delta_{J;J-1} = \varepsilon_{J} - \varepsilon_{J-1}
\end{equation}
are called the hyperfine structure splittings. These values can be measured in modern 
experiments. The coincidence of the experimental and predicted values of $\Delta_{J;J-1}$
can be used to confirm the actual creation of the $e^{+}[{}^3$He$(2^3S)]$ ion. For the 
$e^{+}[{}^3$He$(2^3S)]$ ion we have found the three following splittinings: $\Delta_{1;2} 
\approx$ 5824.3986 $MHz$, $\Delta_{2;0} \approx$ 76466.5308 $MHz$ and $\Delta_{0;1} 
\approx$ 5824.4019 $MHz$. Note that the values $\Delta_{1;2}$ and $\Delta_{0;1}$ almost 
coincide with each other. Formally, it follows from the fact that spin-spin interaction 
between the positron and ${}^3$He nucleus in the $e^{+}[{}^3$He$(2^3S)]$ ion is very small 
(almost negligible), since the corresponding expectation value of the positron-nucleus 
(or positron-helium) delta-function is very small ($\le$ 1.285$\cdot 10^{-6}$ $a.u$.). All
our calculations for the  $e^{+}[$He$(2^3S)]$ ions have been performed with the use of 
KT-variational expansion \cite{KT} of six-dimensional gaussoids in relative four-body 
coordinates $r_{12}, r_{13}, r_{23}, r_{14}, r_{24}, r_{34}$ (for more details see 
\cite{Fro01}). This solves the `mystery' of the hyperfine structure splitting in the 
$e^{+}[{}^3$He$(2^3S)]$ ion. 

In the $e^{+}[{}^4$He$(2^3S)]$ ion we have six spin states which are separated into two
groups: (a) four states with $J = \frac32$ and (b) two states with $J = \frac12$. The
difference between these group of states is $\Delta_{\frac32;\frac12} \approx$ 
55308.6951 $\times \frac32 \approx$ 82963.0427 $MHz$. This frequency corresponds to the 
electron-positron spin-spin interaction. For the both $e^{+}[^3$He$(2^3S)]$ and 
$e^{+}[^4$He$(2^3S)]$ ions the electron-positron spin-spin interaction is the largest 
component of the hyperfine structure splitting. Note again that the both electrons are 
assumed to be in the triplet state, i.e. $S_{-}$ = 1. The hyperfine structure splitting 
in the $e^{+}[{}^4$He$(2^3S)]$ ion is related only with the electron-positron spin-spin 
interaction. For the $e^{+}[{}^3$He$(2^3S)]$ ion the electron-positron spin-spin
interaction is mixed with the two other components of the hyperfine structure splitting: 
with the electron-nuclear spin-spin interaction and with the positron-nuclear spin
interaction. The resulting value of the largest component of the hyperfine structure
splitting in the $e^{+}[{}^3$He$(2^3S)]$ ion decreases to the 76466.5308 $MHz$. We can 
see the two other frequencies (5824.3986 $MHz$ and 5824.4019 $MHz$) which represent the 
hyperfine structure splitting associated mainly with the electron-nuclear spin-spin 
interaction in the $e^{+}[{}^3$He$(2^3S)]$ ion. These values are comparable with the 
analogous splitting in the triplet $2^3S-$state of the ${}^3$He atom: 6740.452154 $MHz$ 
(non-relativistic theory, \cite{Fro02}) and 6739.701177(16) $MHz$ (experiment, 
\cite{Ros}). The difference between the two frequencies 5824.3986 $MHz$ and 5824.4019 
$MHz$ is very small $\le 30$ $kHz$. It can be explained by the contribution from the 
positron-helium spin-spin interaction. In reality, such a small value is very difficult
to measure. However, this value is of great interest, since it can be used as an 
independent measure of the accuracy of modern accurate calculations for the Coulomb 
four-body systems. The hyperfine structure of the excited states in the 
$e^{+}[^3$He$(2^3S)]$ and $e^{+}[^4$He$(2^3S)]$ ions is very similar to the hyperfine 
structure of the ground states in these ions described above.

\newpage 
\begin{table}[tbp]
    \caption{The hyperfine structure and hyperfine structure splitting of the 
             bound state in the $e^{+}[{}^3$He$(2^3S)]$ ion (in $MHz$).}
      \begin{center}
      \begin{tabular}{rrr}
        \hline\hline
      &  $\varepsilon_J$ &  $\Delta_{J;J-1}$ \\
      \hline
$\epsilon_{J=1}$ &  31313.243034 & ------------ \\

$\epsilon_{J=2}$ &  25488.844413 &  5824.39862 \\

$\epsilon_{J=0}$ & -50977.686376 & 76466.53081 \\

$\epsilon_{J=1}$ & -56802.088263 &  5824.40189 \\
       \hline\hline
   \end{tabular}
   \end{center}
   \end{table}
\end{document}